\documentclass[preprint]{aastex} 

\newcommand{\palmsp}{Palm~Vx} 
\newcommand{\degpersec}{\arcdeg\/~s$^{-1}$} 
\newcommand{\minpersec}{\arcmin\/~s$^{-1}$}  
\newcommand{\arcsecpersec}{\arcsec\/~s$^{-1}$}  
\newcommand{\persec}{~s$^{-1}$}  
\newcommand{\newtonmeter}{N~m}
\newcommand{\ampere}{Amp}
\newcommand{\perampere}{\ampere$^{-1}$}
\newcommand{\volt}{Volt}

\begin{document}
\title{The Digital Motion Control System for the Submillimeter Array
  Antennas} \author{T. R. Hunter\altaffilmark{1},
  R. W. Wilson\altaffilmark{2}, R. Kimberk\altaffilmark{2}, P. S.
  Leiker\altaffilmark{2}, N. A. Patel\altaffilmark{2},
  R. Blundell\altaffilmark{2}, R. D. Christensen\altaffilmark{3}, 
  A. R. Diven\altaffilmark{3},  J. Maute\altaffilmark{3}, 
  R. J. Plante\altaffilmark{2}, P. Riddle\altaffilmark{2},
  K. H. Young\altaffilmark{2}}

\altaffiltext{1}{National Radio Astronomy Observatory, 520 Edgemont
  Road, Charlottesville, VA, 22903}

\altaffiltext{2}{Harvard-Smithsonian Center for Astrophysics, 60
Garden Street, Cambridge, MA 02138}

\altaffiltext{3}{Submillimeter Array, 645 North A'ohoku Place, Hilo,
HI 96721} \email{thunter@nrao.edu}


\begin{abstract}

We describe the design and performance of the digital servo and motion
control system for the 6-meter diameter parabolic antennas of the
Submillimeter Array (SMA) on Mauna Kea, Hawaii. The system is divided
into three nested layers operating at a different, appropriate
bandwidth. (1) A rack-mounted, real-time Unix system runs the position
loop which reads the high resolution azimuth and elevation encoders
and sends velocity and acceleration commands at 100~Hz to a
custom-designed servo control board (SCB).  (2) The
microcontroller-based SCB reads the motor axis tachometers and
implements the velocity loop by sending torque commands to the motor
amplifiers at 558~Hz.  (3) The motor amplifiers implement the torque
loop by monitoring and sending current to the three-phase brushless
drive motors at 20~kHz.  The velocity loop uses a traditional
proportional-integral-derivative (PID) control algorithm, while the
position loop uses only a proportional term and implements a command
shaper based on the Gauss error function.  Calibration factors and
software filters are applied to the tachometer feedback prior to the
application of the servo gains in the torque computations.  All of
these parameters are remotely adjustable in software.  The three
layers of the control system monitor each other and are capable of
shutting down the system safely if a failure or anomaly occurs. The
Unix system continuously relays the antenna status to the central
observatory computer via reflective memory.  In each antenna, a
\palmsp\/ hand controller displays complete system status and allows
full local control of the drives in an intuitive touchscreen user
interface.  The hand controller can also be connected outside the
cabin, providing a major convenience during the frequent
reconfigurations of the interferometer.  Excellent tracking
performance ($\sim 0.3\arcsec$ rms) is achieved with this system. It
has been in reliable operation on 8 antennas for over 10 years and has
required minimal maintenance.

\end{abstract}

\section{Introduction}

The successful operation of tracking interferometers for astronomy
demands a safe, reliable, and accurate servo system to control the
motion of the primary reflector and the supporting mount of each
telescope.  For radio interferometers, the pointing and tracking
requirements of the individual telescopes depend upon the reflector
diameter and the observing frequency.  The operational safety
requirements include the protection of the mechanical structure of the
telescope and the protection of observatory staff and visitors to the
site.  Safety is an essential concern for facilities that support
remote observing, particularly in areas where public access is
allowed.

The Submillimeter Array (SMA)\footnote{The Submillimeter Array is a
  joint project between the Smithsonian Astrophysical Observatory
  (SAO) and the Academia Sinica Institute of Astronomy and
  Astrophysics (ASIAA) and is funded by the Smithsonian Institution
  and the Academia Sinica.}  consists of eight 6~m diameter antennas
arranged among 24 concrete pads near the summit of Mauna Kea, Hawaii
\citep{Blundell07,Ho04,Moran98}.  The various specifications for the
SMA place stringent requirements on the antenna servo for tracking
accuracy, slew rate, and settling time.  For interferometers, tracking
errors result in two deleterious effects: (1) loss of signal from
sources at the phase center due to the finite extent of the
single-dish primary beam response, and (2) improper weighting of
radiation over the synthesis area leading to loss of image fidelity,
especially for mosaicked observations of extended objects
\citep{Bhatnagar08}.  The specified frequency range of the receivers
covers 180-900 GHz \citep{Blundell05,Hunter05}, resulting in a primary
half-power beam width that ranges from $\theta_{\rm beam}=66$ to 13
arcseconds.  A desirable goal for pointing accuracy is to maintain
$\Delta\theta < (\theta_{\rm beam}/20)$, which will hold the total
power fluctuations below 0.5\% for a source at the center of the
primary beam \citep{Napier89}.  For the SMA to meet this criterion at
the highest operating frequency, the pointing and tracking errors must
be kept below 0.7\arcsec.

The ability to perform ``fast switching'' between the target source
and a gain calibrator is also desirable for optimal gain calibration
of the interferometer \citep{Holdaway1995}, particularly when
observing in the highest frequency bands \citep{Hunter07}.  In
practical terms, the number of quasars brighter than 1~Jy at 230~GHz
and accessible to the SMA is usually about 50 and they are coarsely
distributed over the observable sky of $\sim36000$ square degrees
\citep{smacatalog, carmacatalog}. Thus, the typical angle between a
target source and the nearest calibrator is $\sim$15\arcdeg\/,
although the recent improvements to the SMA bandwidth and receiver
performance \citep{Tong12} will tend to push this value lower as
fainter calibrators can be used.  In any case, the power spectrum of
atmospheric path length variations versus time typically rises with an
exponent of $0.7$ \citep{Masson94, Kimberk12}, starting from the
shortest time scale -- the antenna diameter divided by the wind speed
(about 1 second for the SMA on Mauna Kea) -- up to a characteristic
timescale (100-200 sec for Mauna Kea) beyond which the spectrum of the
variations flatten.  This timescale is often called the ``corner
time'' and corresponds roughly to the time required for the turbulent
phase screen to cross the interferometer.  If left uncorrected, these
variations will cause decorrelation of the astronomical measurement.
This atmospheric decorrelation can be reduced if the calibrator and
target source are close together on the sky and are observed
periodically with a cadence less than the corner time.  Thus, slew
speeds of a few degrees per second and rapid settling times (of 1-2
seconds) are necessary to maintain a high duty cycle in observing the
target source.


There are remarkably few papers in the astronomical literature that
describe the design of high performance telescope servo systems and
encompass all the practical aspects involved from high-level software
to low-level hardware.  One of the most complete examples is the
description of the digital servo developed for the Onsala 20~m
telescope \citep{Olberg95}.  The goal of this current paper is to
present the detailed design and performance of a modern digital servo
and motion control system that has been implemented on the SMA
antennas, including a description of the extensive safety features and
the novel portable controller.  Not included is any discussion of the
control system of the subreflector mirror which is based on a
commercially available controller \citep{Cheimets94}, nor is the
calibration of the telescope pointing models treated, as it is
described in \citet{Patel04}.  Although the mechanical components of
the SMA antennas were constructed by subcontractors, the SMA partners
(SAO and ASIAA) assembled the antennas which provided an opportunity
for the in-house development of a precise servo system to meet the
scientific needs of the telescope.  The scientific and technical staff
of the SAO submillimeter receiver lab designed, fabricated, and tested
the servo in 1998-1999.  It was first commissioned and tested on an
SMA antenna at the temporary SAO site at Haystack Observatory in late
1999 and early 2000.  Many improvements and conveniences to the
software were added over the subsequent two years based on experience
from operations at the Mauna Kea site.  The paper begins by describing
the fundamentals of antenna servo control in \S\ref{theory}, including
some relevant information on the SMA antenna structure and its open
loop response.  The hardware and software design of the servo system
is presented in \S\ref{design} and the \palmsp\/ portable controller
is described in \S\ref{interface}.  The key performance measurements
are given in \S\ref{performance}.

\section{Antenna servo fundamentals}
\label{theory}

\subsection{Basic equations}

The fundamentals of servo systems for accurate dish antenna
positioning were first presented by \citet{Lozier63} for hydraulic
drives.  Regardless of the type of drive, motors are typically
connected to antennas through gearing, which effectively reduces the
moment of inertia of the antenna ($J_a$) as seen at the motor shaft.
The equivalent or ``reflected'' inertia ($J_{eq}$) is given by:

\begin{equation}
J_{eq} = R^2 J_{a} = \biggl(\frac{N_{1}}{N_{2}}\biggr)^2 J_{a} 
\end{equation}

where $R$ is the gear ratio, for example in a two-gear system of
$N_{1}$ and $N_{2}$ teeth \citep{Klafter89}.  Alternatively, a similar
reduction in antenna inertia can be achieved with a leadscrew or ball
screw drive, which converts rotary motion of a shaft into linear
motion of a mass ($M$) through its coupling ratio or pitch ($P$) 
in units of turns per distance traveled.  In this case, the equivalent 
inertia is given by:
 
\begin{equation} 
J_{\rm eq} = \frac{M}{(2\pi P)^2}.
\end{equation}

Because only a limited range of motion ($<90\arcdeg$) is typically
required, leadscrews are sometimes employed to drive antennas in
elevation, for example the Leighton antennas of the Caltech
Submillimeter Observatory \citep{Woody94} and the Combined Array for
Millimeter Astronomy \citep{Woody04}.

The moments of inertia of the antenna ($J_{a}$) and the motor
($J_{m}$) for one axis of motion, combined with the maximum desired
values for the angular velocity ($\dot{\theta_a}$) and angular
acceleration of the antenna ($\ddot{\theta_a}$) for that axis, will
determine the levels of motor torque ($T_m$) that must be generated to
meet the system goals.  The fundamental relation between these
quantities and the motor torque constant ($\kappa_T$), the applied
armature current ($I$), the Coulomb and viscous bearing friction
coefficients ($C$ and $B$), which may both be a function of their
respective axis angle ($\theta$), and the spring constant of the
system ($k$) is given by the second order differential equation for a
moment of inertia-damper-spring system running open-loop\footnote{Not
  represented in the Coulomb friction term in equation~\ref{torque} is
  the property of static friction (stiction) which is an increase in
  friction at zero velocity.}  \citep[see e.g.][]{VandeVegte93}:

\begin{equation}
\label{torque}
T_m = \kappa_T I = 
\overbrace{(J_m+R^2J_a)}^{\rm inertia}\ddot\theta_m + 
{\overbrace{(B_{m,\theta_m}+R^2B_{a,\theta_a})\dot{\theta_m}}^{\rm viscous~friction}}+ 
\overbrace{(C_{m,\theta_m}+RC_{a,\theta_a})}^{\rm Coulomb~friction} + 
\overbrace{k(\theta_m-R\theta_a)}^{\rm spring} 
+ RT_{\rm wind}
\end{equation}

The torque applied by the wind to an antenna axis ($T_{\rm wind}$) is
a complicated function proportional to the air density, the wind
velocity squared, and an integral over the area of the antenna
structure projected toward the wind times an effective radius from the
axis and a drag coefficient \citep[see e.g.][]{Levy96,Campbell04}.
The torque delivered to the antenna axis ($T_a$) can be written as a
sum of the antenna frictional terms and the remaining torque available
to accelerate the antenna load ($J_a\ddot{\theta_a}$):

\begin{equation}
T_a \equiv C_{a,\theta_a} + T_{\rm wind} + B_{a,\theta_a}\dot{\theta_a} + J_a\ddot\theta_a = T_m - \frac{J_m}{R^2} -  
\frac{B_{m,\theta_m}\dot{\theta_a}}{R^2} - 
\frac{C_{m,\theta_m}}{R} - 
k\biggl(\frac{\theta_m}{R}-\theta_a\biggr)
\end{equation}

The torque required to maintain even a constant angular velocity will
vary with time and angular position of the telescope due to any number
of minor disturbances including variable friction in the gear dynamics
and bearings, and wind forces.  Thus, the goal of a servo system is to
continuously observe the dynamic state of the antenna by reading
feedback instruments, computing the following error with respect to
the desired or expected trajectory, and adjusting the applied torque
in order to minimize this error.  Typically, the resonant frequencies
of the mechanical structure impose a practical limit on the maximum
value of $\ddot{\theta}$ that one can apply.  In other words, to avoid
exciting the structure, the servo bandwidth must be limited to less
than the lowest resonant frequency, or employ notch filters to mask
the response.  Antenna servo design draws upon the general techniques
of feedback control systems which are described in various engineering
textbooks \citep[see e.g.][]{Franklin05}.  While there has been recent
research on alternative servo algorithms for large radio telescopes
\citep{Smith2010,Zhu07,Gawronski05}, the conventional
proportional-integral-derivative (PID) control algorithm has been the
most traditional approach.  Fundamentally, a PID-based controller
periodically computes the difference between the measured position and
velocity and their corresponding target values. The P term is based on
the present error, the I term is based on the accumulation of errors
in the (recent) past, and the D term is based on how rapidly the error
is currently changing.  These three terms are typically combined into
a weighted sum using individual gain factors before being applied to
the motor amplifier as the new desired torque value.

\subsection{SMA antenna properties}

\begin{figure}[h!]
\includegraphics[angle=0,width=6.25in]{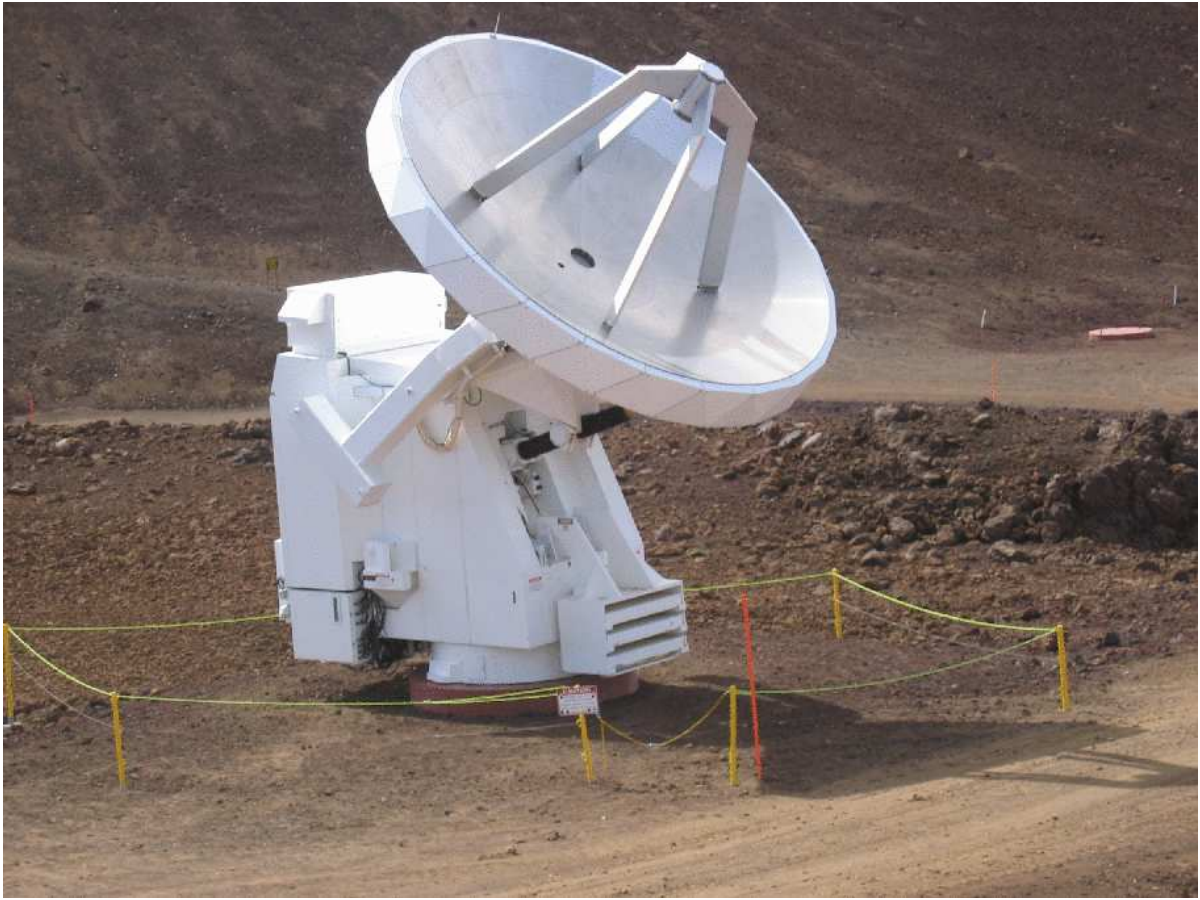} 
\caption{Photograph of an SMA antenna on Mauna Kea.  The azimuth
  counterweights are located at the bottom front of the mount and the
  elevation counterweights are the inverted T-shaped pieces behind the
  dish.  The elevation ballscrew protrudes from the cabin and is
  covered by the black cylindrical bellows.
\label{antenna}}
\end{figure}

\begin{figure}[h]
\includegraphics[angle=-90,width=6.5in]{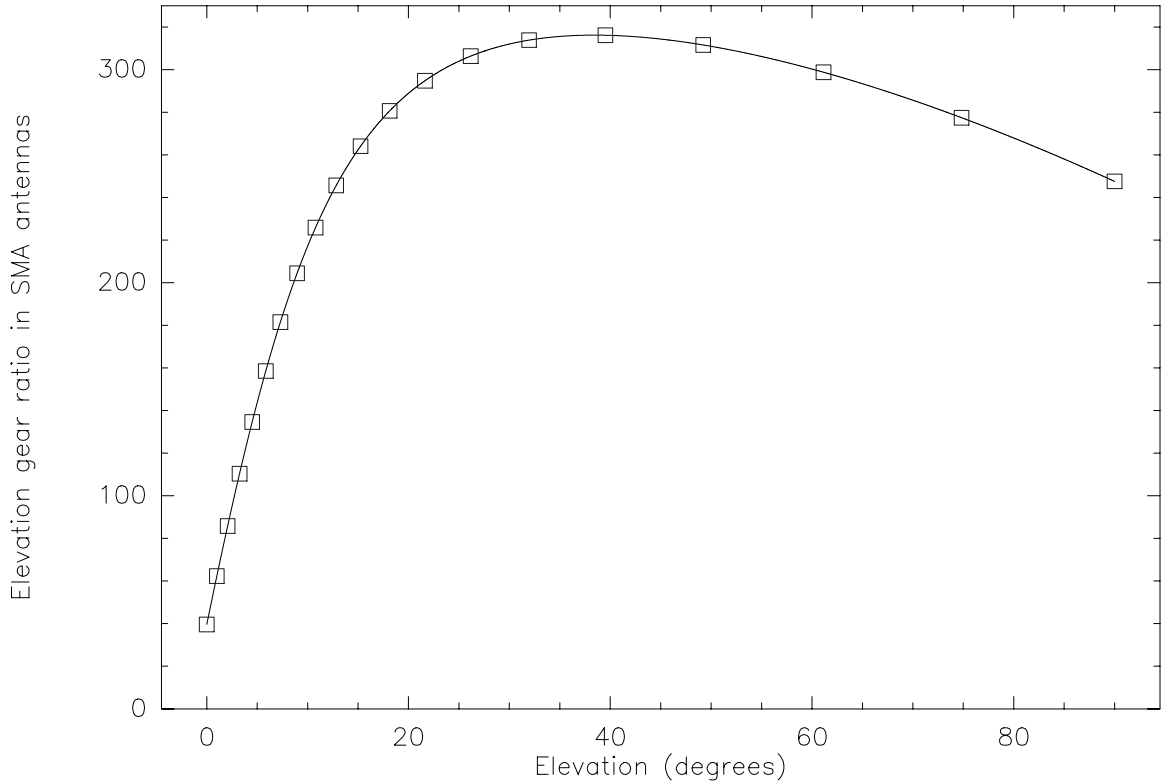}
\caption{The gear ratio of the SMA elevation ballscrew drive as a function of 
antenna elevation, given in degrees above the horizon. 
\label{elevationGear}}
\end{figure}

The SMA antennas (Figure~\ref{antenna}) consist of a 6~m diameter
parabolic reflector comprised of aluminum panels attached to a backup
structure of carbon fiber tubes and steel nodes sitting atop a stiff
steel mount which houses the receiver system in a temperature
controlled environment.  The antennas are driven along azimuth and
elevation axes using brushless three-phase motors with a peak torque
of 1085~\newtonmeter\/, similar to those used by the Cosmic Background
Imager \citep{Padin02}.  In azimuth, there are two motors meshed to a
2.14~m-diameter bull gear at a tooth ratio of 422:30 (14.0667:1).  The
bull gear remains stationary with respect to the pad as the rotating
portion of the antenna mount drives around it.  In elevation, a single
motor drives a ballscrew whose effective gear ratio varies with
elevation from $\sim$315 at 45\arcdeg\/ dropping to $\sim$250 at
15\arcdeg\/ and 87\arcdeg\/ (Fig~\ref{elevationGear}).  The ballscrew
nut moves away from the motor as the telescope moves upward in
elevation.  The total range of motion between the hardstops is
-178\arcdeg\/ to +357\arcdeg\/ in azimuth and 6.7\arcdeg\/ to
89\arcdeg\/ in elevation.  The relatively small gear ratio in azimuth
means that the reflected moment of inertia (including counterweights)
is dominated by the antenna structure (61450~\newtonmeter\/ at zenith,
311~\newtonmeter\/ reflected), whereas in elevation the inertia
contributed by the motor and the directly-driven lead screw
(1.39~\newtonmeter\/) is of similar magnitude to that of the antenna
(58270~\newtonmeter\/, 0.95~\newtonmeter\/ reflected).  Thus, it is
clear that azimuth will be the more challenging axis to control
smoothly as the antenna dynamics will be more influential.

\subsection{Open loop response}

When designing a servo, it is crucial to first understand the
frequency response of the system to the applied torque.  Finite
element models (FEM) of the SMA antenna indicate that the lowest
natural frequency of the telescope assembly is a bending mode of the
mount at 14~Hz \citep{Raffin91}. Additional bending and torsional
modes of the reflector, backup structure, and quadrupods occur at
frequencies above 16~Hz.  These predictions were tested after
construction of the prototype antennas at Haystack Observatory in the
following manner.  One azimuth drive motor was excited with sinusoids
at a sequence of frequencies and the response of the high resolution
encoder was recorded in order to generate a Bode plot of the open loop
response.  Significant resonances at 13 and 19~Hz were found, which
differ somewhat from the FEM result, along with an additional weaker
feature near 8~Hz.  In a similar experiment, one azimuth motor was
excited with sinusoids while the response of the tachometer on the
other azimuth motor was recorded.  A resonance in the elevation
counterweights was found which explained the lowest frequency feature.
This is not surprising since the FEM was made for the original design
of the SMA antenna which was unbalanced in both axes and planned to
use "electronic counterweights" instead of physical counterweights.
Clearly these resonances will need to be avoided by the servo
controller.

\section{Servo System Design}
\label{design}

\subsection{Conceptual design}

\begin{figure}[ht]
\includegraphics[width=6.5in]{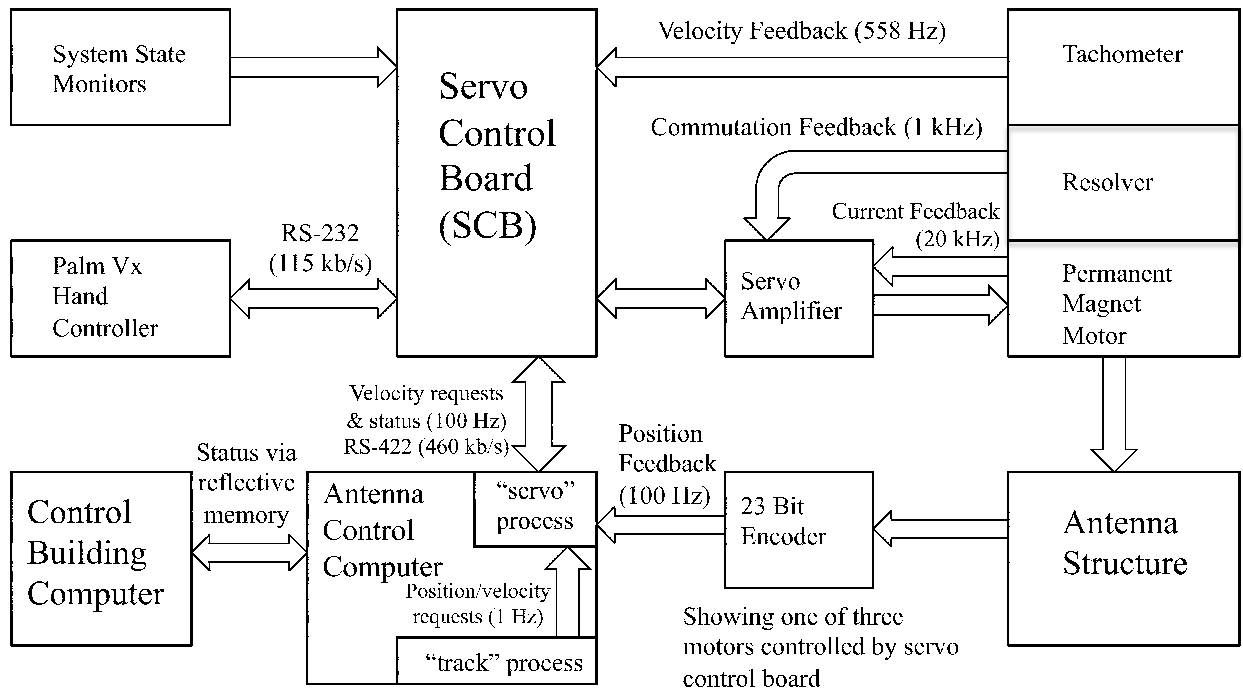}
\caption{Diagram of the servo control system for the SMA antennas.
  The feedback occurs at the following rates: position: 100~Hz,
  velocity: 558~Hz, motor commutation: 1~kHz, and motor current:
  20~kHz.
\label{diagram}} 
\end{figure}

Initial attempts to drive the prototype SMA antennas using a
commercial servo controller yielded less than optimal results and had
a rather arcane command interface.  It also suffered from considerable
stiction in the elevation drive screw at low elevation angles because
of the lack of counterweights and poor mechanical advantage at low
elevation angles.  As a result, it was decided to add physical
counterweights and to create a reliable digital servo control system
designed around a well-established microcontroller placed on a custom
integrated circuit board comprised of standard components and
interfaces, and programmed in the widely-adopted C programming
language.  This design led to an effective division of the PID control
concept into three nested layers, each implemented in a different
processor, and each running an independent PID servo loop.  While it
may seem complicated, the design leads to a natural division of labor
between the motor amplifier circuitry, a rugged embedded system, and a
networked Unix computer.

A basic diagram of the design is shown in Figure~\ref{diagram}.
Forming the bottom layer, the motor amplifiers execute the highest
bandwidth ($\sim$~20~kHz) servo loop by monitoring and controlling the
current delivered to the motors.  The middle layer is implemented by a
custom-designed servo control board (SCB) containing a microcontroller
which closes the velocity servo loop around the motors by reading the
tachometers, calculating new torque commands and sending them as
voltages to the motor amplifiers at 558~Hz.  This loop is a full PID
loop with relatively high integral gain so it behaves somewhat like a
position control on the motor shaft.  (The derivative term is zero in
azimuth and not very important in elevation.)  In the case of torque
overload (that is, when the computed torque request would exceed the
maximum output value), the integral term is constrained to a smaller
(or zero) value in order to constrain the delivered request to be
within the controlled range.

The highest servo layer executes on each antenna control computer
(ACC), a diskless, network-booted, VMEbus Power PC running a flavor
of real-time Unix (LynxOS).  A server process on this machine
administers the position servo loop by receiving position and velocity
requests once per second, reading the position encoders, and computing
and sending velocity commands to the SCB at 100~Hz.  This top level
loop is a simple position loop with relatively low gain to avoid
exciting structural resonances.  It compensates for the low gain by
having accurate velocity feed forward to the relatively stiff motor
velocity loop.  The 100~Hz timing is maintained via programmed
interrupts received from a synchronized generator card on the VMEbus.

Among the three layers, there is redundant monitoring of control
parameters, as the three layers monitor each other and are capable of
shutting down the system if any measured quantities become out of the
nominal range.  Conceptually, the three layers also divide the task of
programming in a natural progression, with the Power PC being the
easiest to program as it requires no knowledge of the underlying
hardware, while at the other end, the optimization of the motor
amplifier circuitry requires advanced electrical engineering skills.
Although this digital control system could easily be used to implement
and test alternative control algorithms, such as
linear-quadratic-Gaussian (LQG) and H-infinity \citep{Gawronski01},
the performance of this more traditional PID system meets the required
specifications, after a couple of digital tweaks such as resetting the
integrator upon overload and using an optimal strategy for torque bias
(\S\ref{bias}).

\subsection{Hardware components}

\subsubsection{Motors and motor amplifiers}

Each motor is a 32 pole brushless, permanent magnet rotor, torquer.
The stator electromagnets are driven by three-phase power emitted by a
pulse width modulated (PWM) current controller
(7~\ampere\/~\volt$^{-1}$) operated at $\sim 20$~kHz and located on
the motor amplifier circuit boards (manufactured by Glentek).  Motor
current feedback is supplied by Hall effect transducers.  The
torque constant of the motors is $\kappa_T =
16.3$~\newtonmeter\/~\ampere$^{-1}$ with a maximum acceleration of
3070~rad~s$^{-2}$.  Considering the gear ratios, the motor-amplifer
system produces 0.041 rad~s$^{-2}$~\perampere\/ at the azimuth motor
and 23.6 rad~s$^{-2}$~\ampere$^{-1}$ at the elevation motor.  In terms
of dish motion, these values equate to
0.17\arcdeg\/~s$^{-2}$~\perampere\/ in azimuth and
4.5\arcdeg\/~s$^{-2}$~\perampere\/ in elevation.

A small resolver is attached to the shaft of each motor and is excited
at 1 kHz. Its feedback is used in apportioning the drive current to
the three motor windings so that the resulting magnetic field exerts
the desired torque on the permanent magnet rotor. The motors can be
commutated at any one of the six possible motor phase combinations.
Due to the low inductance of the torquer (6~mH), series inductors were
added to each of the three power wires in order to increase the time
constant for the current rise and thereby reduce the level of
electrical noise emitted into the electronic environment of the
antenna.  As a safety feature, the maximum velocities were set in
hardware to approximately 6\degpersec\/ (azimuth) and 3\degpersec\/
(elevation) by choosing the appropriate resistors and capacitors for
the components associated with the AD2S83 resolver-to-digital
converter integrated circuit.  This device demodulates the resolver
sine/cosine output and provides a 16 bit output for the commutation
circuit.  The azimuth value was chosen to prevent the dish from
breaking loose from the mount during a rapid deceleration from this
velocity when encountering the hardstop, and the elevation value was
chosen so that the stopping impact would not exceed the maximum
dissipation of the rubber hardstops. The servo system in turn limits
the maximum controlled slew velocities to 4\degpersec\/ and
2\degpersec\/, respectively.  The amplifiers have a dynamic brake
feature which is disabled during normal operation but will engage in a
commanded or emergency shutdown.

\subsubsection{Microcontroller}

\begin{figure}[hb!]
\includegraphics[width=6.0in]{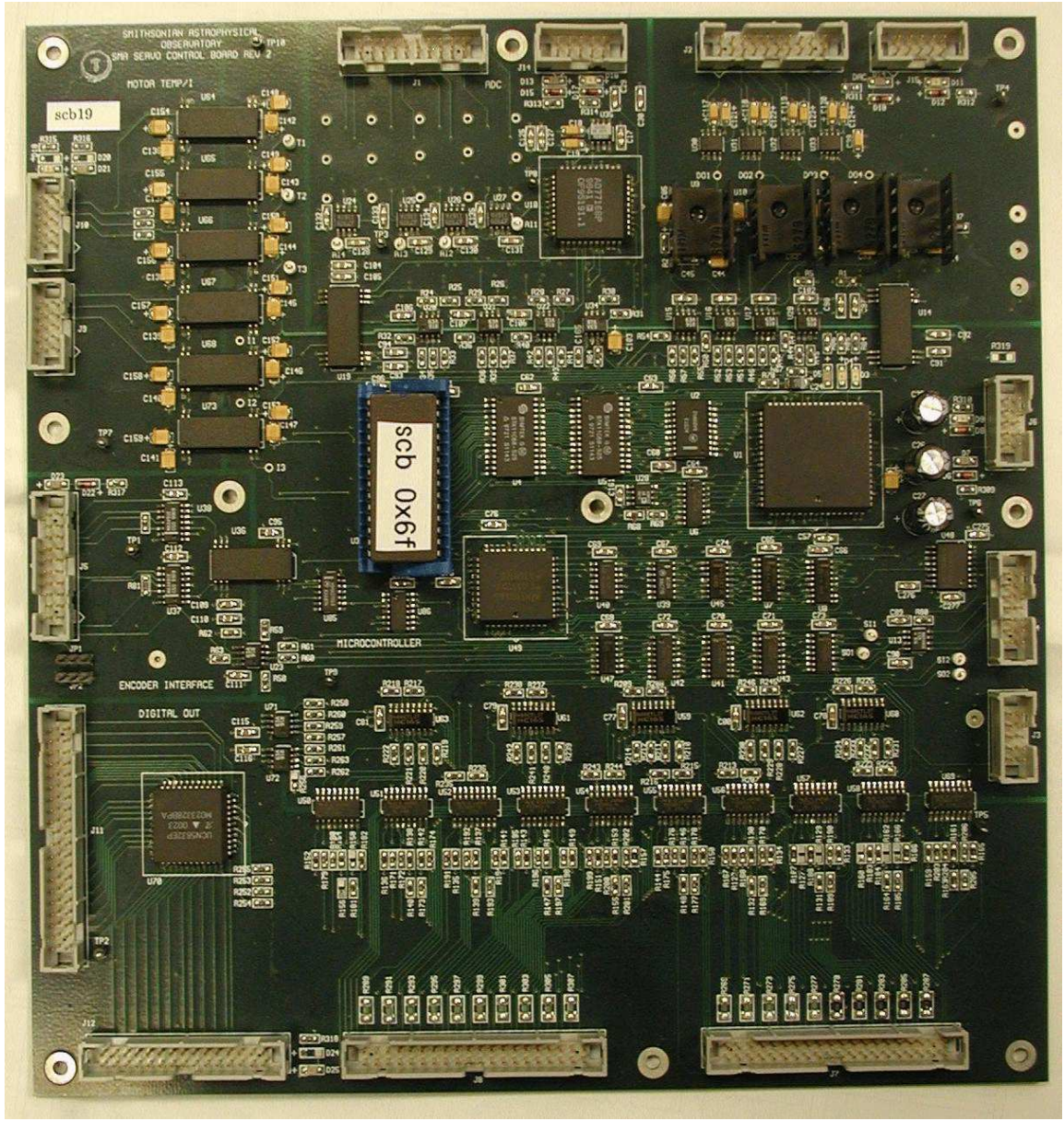} 
\caption{Photograph of the 24~cm servo control board (SCB) which is
  installed in each SMA antenna.  The Intel microcontroller is the large
  integrated circuit located to the right of the center, and the EPROM 
  containing the control code is the chip to the left of center with 
  the version label attached and installed in a zero-insertion force 
  socket.
\label{scb}} 
\end{figure}

The SCB is built on a $24 \times\/ 24$~cm four-layer printed circuit
board (see Figure~\ref{scb}).  The processor at the heart of the SCB
is an Intel 80C196KC 16-bit microcontroller running at 20~MHz on an
8-bit bus.  This is the same controller used to operate the digital
phase lock loop and Gunn oscillator and attenuator motors on the
front-end first local oscillators \citep{Hunter11,Hunter02}.  A C
language cross-compiler (Tasking, Inc.) is used to generate the
executable code.  The microcontroller manages a multitude of hardware
devices:
\begin{itemize}
\item 40 digital inputs: to read limit switches and fault indicators
\item 32 digital outputs: to enable, disable, or reset the motor amplifiers,
apply or release the mechanical or dynamic brakes, and to drive the 
environmental door that protects the receiver cabin from the weather
\item  a four channel 22-bit $\pm2.5$~V analog to digital converter (ADC): 
to read the motor tachometers
\item a four channel 16-bit $\pm5$~V digital to analog
converter (DAC): to provide torque commands to the motor amplifiers
via balanced differential drivers
\item an RS-422 serial port: to communicate with the ACC (460~kbps)
\item an RS-232 serial port: to communicate with the \palmsp\/ hand 
controller (115~kbps) 
\item synchronous serial ports: to read the low-resolution ''coarse'' 
encoders on the azimuth and elevation axes.  
\item a 32-LED indicator panel with serial interface: to display system 
status to personnel working in the antenna cabin
\end{itemize}
Each section of the board employs independent ground planes and power
supplies, with optical isolation on all I/O channels.  There are
convenient test points installed to allow easy debugging of the analog
signals either arriving at, or generated by, the SCB.  The
microcontroller program runs from an erasable programmable read only
memory (EPROM).  During the development phase of the SCB, over a dozen
different simple programs were devised and written to individual
EPROMs to test specific hardware devices on the board and which
displayed the test results on the LED status panel.  For the third
(and final) revision of the circuit board, twenty boards were
constructed.

\subsubsection{Feedback devices}
\label{feedback}
\subsubsubsection{Antenna velocity}

Motor velocity feedback is provided by brushed tachometers that are
installed on each motor axis and have a gain of 12~V/(rad~s$^{-1}$),
which equates to 2.95~V/(deg s$^{-1}$) in azimuth and 63~V/(deg
s$^{-1}$) in elevation.  The tachometer signals are divided by
resistor networks on the SCB ($\div 4.734$ in azimuth and $\div 53.22$
in elevation) and sampled by a 4-channel Sigma-Delta 22-bit ADC.
Following digitization, the SCB applies the empirically derived
tachometer calibration factors that are stored in memory so that the
velocity feed-forward will be accurately applied.  Whenever the servo
is disabled, the ADC readings are still continuously collected and
used to automatically define the zero offset value.  The 10-bit ADC
onboard the microcontroller is used to monitor the temperature and
current of the three motors via an eight-input multiplexer.

\subsubsubsection{Antenna position}

Antenna axis position feedback is provided by two sets of encoders:
``coarse'' encoders that provide absolute antenna position to the SCB
and ``fine'' encoders which provide high resolution absolute position
to the ACC.  The coarse encoders (manufactured by Stegmann) are 24-bit
devices with 4096 counts per turn and 4096 turns connected to the SCB
via a synchronous serial interface.  They are connected to the antenna
axes by gears and have a resolution of about 0.0052\arcdeg\/ in
azimuth and 0.023\arcdeg\/ in elevation at the antenna axes.  The
calibrated offset values corresponding to zero azimuth and zero
elevation are stored on the SCB in non-volatile static random access
memory (NVSRAM) backed up by an electrically erasable read only memory
(EEPROM).  These encoders are used by the SCB to enforce ''soft''
limits in order to avoid engaging the hardware limit switches during
normal operations, which would shut down the system and require
operator intervention.  The nominal values of the soft limits are
-171\arcdeg\/ to +349\arcdeg\/ in azimuth and 14\arcdeg\/ to
87.5\arcdeg\/ in elevation.  The SCB also repeatedly sums the
tachometer signals for 100 servo cycles and compares the result to the
change in the corresponding coarse encoder as a safety check that both
devices are working properly.

A high resolution encoder (manufactured by Heidenhain) is directly
driven by each axis.  They are read using Heidenhain's EnDat
\textsuperscript{\texttrademark}\/ protocol by the position loop on
the ACC, which stores the corresponding zero values in a configuration
file.  The original encoders had 23 bits (for a resolution of
0.154\arcsec\/) but a few antennas have a newer model of the same
geometry with 27 bits.

\subsubsubsection{Motor faults and limit switches}

A large number of digital inputs bring system information to the SCB.
Each of the three motor amplifiers produce a fault signal which is
monitored by the SCB.  Two pairs of limit switches are installed near
the end of travel of each axis: a ``pre-limit'' which triggers a
software servo shutdown, and a ``final'' limit which disengages the
power relay to the motor amplifiers for that axis.  The open collector
arrangement guarantees that a broken wire always causes the limit
switch to appear as if it was engaged.  The SCB can be commanded to
temporarily override the soft limits and/or the pre-limits (for a
fixed time interval) to allow manual escape of a pre-limit via the
\palmsp\/ hand controller.  In order to escape an engaged final limit
switch, there is a momentary button switch in the cabin which will
allow the power relay to be energized while the button is depressed.
In addition, there is a limit switch that monitors the orientation of
the azimuth rocker arm, which deploys the azimuth hardstop in one of
two positions depending on which azimuth wrap the antenna is currently
operating (the normal azimuth range is -176\arcdeg\/ to
+353\arcdeg\/). Finally, there are two limit switches that indicate
the position of the environmental door (fully closed or fully open)
which is located just above the antenna elevation axis and can be
commanded to seal the cabin from the outdoor weather.  In this case,
it blocks the optical path from the subreflector to mirror \#3, a flat
mirror which rotates on the elevation axis and illuminates the fixed
receiver optics \citep{Paine94}. Hence, it is often called the ``M3
door''.

\subsubsubsection{Pad identification} 

Five digital inputs are used to sense the identification code for the
antenna pad (the pad ID).  This code is established by resistors in a
cable terminator attached to each pad, and is relayed back from the
SCB to the ACC and the SMA control system.  The SCB uses this code to
recognize when it is on the transporter and adjusts the servo gains to
the appropriate preset values automatically.  It also uses this code
to recognize when it is occupying a position on the inner ring of
eight pads where a collision between antennas can be possible if the
array is in the ``subcompact'' configuration.  In this case, the SCB
will read an additional limit switch which effectively increases the
low elevation software limit to 32\arcdeg\/.  For additional safety, a
physical wedge is also installed on the antennas in this configuration
to raise the elevation hard stop to $\sim30$\arcdeg\/, which is
sufficient to preclude any collision.  There is a mechanism in
software to use a fake pad ID if the readback code corresponds to an
invalid pad number.  Following the commissioning phase of the SMA,
this feature has been used only rarely (and temporarily) in the case
that the cable terminator is damaged or left unconnected after a
reconfiguration.

\subsubsubsection{Emergency stops}

An emergency stop (E-stop) circuit is implemented in each antenna
which cuts power to the motor amplifiers.  It is comprised of a
sequence of buttons both inside and outside the antenna cabin, on the
\palmsp\/ hand controller, and in the central control building via a
fiber optic link.  In addition, the ACC can engage the E-stop via
a digital output bit that controls a switch in the circuit.  One
digital input on the SCB monitors the E-stop circuit status.

\subsubsubsection{Coolant flow and brakes} 

Because all three motors are liquid cooled, there is a coolant flow
sensor located along the coolant flow tube which indicates whether
there is sufficient flow present and it is monitored by the SCB.  The
azimuth brakes are released by air pressure, and the pressure in the
air ballast tank is monitored by a gauge which produces a low pressure
signal connected to the SCB.  The release status of the elevation
brake is also monitored.

\subsubsubsection{Status display panels}

Continuous feedback to personnel in the antenna cabin is provided by
the SCB via a 15~cm panel of 32 color-coded LED indicators which
indicate the status of the servo and all the associated limit switches
and possible error conditions.  In normal active operation, all
indicators illuminated are green, while error conditions are shown by
red indicators.  In a similar manner, the commercial motor amplifier
cards have 13 color-coded indicators that indicate either normal
operation or which reason(s) has triggered the fault.

\subsection{Software components}

\subsubsection{SCB microcontroller software}

The 80C196KC microcontroller executes an interrupt-driven,
scheduler-based program written in C and assembled into the 64~kB EPROM
with the program variables (and some functions) stored in two banks
of 8~kB of NVSRAM/EEPROM.

\subsubsubsection{Scheduler}

The main loop of the EPROM program is a subroutine scheduler.
Whenever the program is not servicing an interrupt, the scheduler
continuously works its way down through a possible list of 11 jobs in
order of descending priority.  The list of jobs along with a brief
summary is given below:
\begin{enumerate}
\item FASTSERIAL: read the encoders and digital inputs; schedule the SAFETY and 
WATCHDOG jobs
\item SYSTEM: sequence the startup/shutdown (motor amplifier power and coolant 
pump)
\item AZ.SEQUENCE: enable/disable the azimuth servo and/or braking 
\item EL.SEQUENCE: enable/disable the elevation servo and/or braking 
\item ACC.REQUEST: process the serial packet commands from the ACC
\item SAFETY: check the digital inputs, encoder range, and ADC status; schedule the SYSTEM, 
AZ.SEQUENCE, or EL.SEQUENCE jobs as needed
\item SLOWSERIAL: update the digital outputs and the LED status board
\item M3: operate the ``mirror \#3'' environmental cabin door
\item HORN: sound the external horn prior to a significant slew
\item PALMPADDLE: send the servo status variables to (and/or process 
commands from the \palmsp)
\item WATCHDOG: clear the timer if all is well and ACC packets are being 
received
\end{enumerate}

Each job has an indicator as to whether it has been scheduled and
needs to be executed.  For example, when the FASTSERIAL job completes,
the program checks if the SYSTEM job has been scheduled.  If so, it
executes that job, otherwise it next checks AZ.SEQUENCE, and so on.
The activity of the scheduler is paced by two interrupt service
routines that run periodically and always take precedence over the 11
jobs above: the ADC ``data ready'' interrupt and the ``incoming serial
data'' interrupt.  The ``data ready'' pin of the 22-bit ADC
periodically drives the main interrupt of the microcontroller at a
rate of 558~Hz, triggering the ADC interrupt service routine which
reads the ADC, computes new torque commands, and schedules calls to
SLOWSERIAL and FASTSERIAL.  Similarly, the arrival of an RS-422 serial
packet at the universal asynchronous receiver/transmitter (UART) chip
on the SCB triggers the serial interrupt service routine which decodes
the packet and schedules ACC.REQUEST. Under normal operation, the
velocity requests arrive as serial packets at 100~Hz from the
command-shaping, position loop servo process running independently on
the ACC.  Each packet contains: (1) a start byte beginning with 0xc0
and encoded with the packet type (of which there are 64), (2) the
content bytes, (3) a two byte value resulting from a cyclic redundancy
check calculated over the content bytes, and (4) a new line character
as the stop byte. The content bytes (up to 32) are encoded such that
their sign bits are stripped and assembled into an additional byte
included after every sixth content byte.  The SAFETY job, which
performs a variety of telescope safety checks such as reading limit
switches, gets scheduled (by the FASTSERIAL job) on every other
interrupt cycle (279~Hz) and can trigger shutdowns accordingly.
Similarly scheduled, the WATCHDOG job performs the routine servicing
of a 6 millisecond watchdog timer; thus if the microcontroller CPU is
ever overcome by the jobs to be run, it will be reset very quickly,
and its default outputs upon restart will trigger an immediate
shutdown of the servo.

The microcontroller provides register space for the storage of
important variables.  This space is used to record the amount of spare
time remaining upon completion of all the tasks in each interrupt
service execution.  It also contains two 32-bit fault words: the
current list of faults and the list of faults present upon the last
execution of the non-maskable interrupt handler, which will run only
in the event that the microcontroller is reset or powered off.  The
stored faults will then only be cleared after they are read by the
ACC.  Because the register variables are unspecified at power up, a
mirror copy of each is kept in SRAM and copied to the registers at
system startup.  This includes a copy of the low resolution encoder
positions, which is useful in the event that an encoder needs to be
temporarily removed, as the final received antenna position will be
retained and continue to be displayed to the rest of the upstream
system after a power cycle, up until the encoder is able to be
reconnected.

\subsubsubsection{Velocity servo loop parameters}

The velocity servo loops receive updated velocity and acceleration
commands at 100~Hz from the ''servo'' process on the ACC.  The
acceleration value is used to smoothly interpolate the velocity
commands between updates to reduce noise and, probably, mechanical
wear.  Two digital software lag filters centered at 12~Hz and 18~Hz
are applied to the azimuth tachometer feedback signal to prevent
amplification of azimuth resonances.  The gains and filter parameters
are set in software and can be remotely modified.  This capability is
most critical during the commissioning and servo tuning phase as it is
not possible to pre-determine optimal values for all the servo
parameters \citep{Gawronski07}.  The optimal gains will provide a
critically damped step response which does not contain spectral
components that excite the structure.  In order to arrive at the
optimal gains, the gains were adjusted manually to obtain fast
acquisition after a slew, and smooth tracking at various speeds.  The
general approach was to start with low gain and increase it until
overshoot or oscillatory behavior started, increasing the damping as
helpful, but eventually backing off to achieve the fastest settling
time to within an arcsecond of the new trajectory after a slew, over a
variety of cases. The tools described in \S\ref{accsoftware} were used
to drive the system and collect performance data during this process.

The ability to modify the gains in software is also important for long
term operations of antennas that are part of reconfigurable
interferometers like the SMA, as the servo gains will be quite
different when the antenna is carried by a transporter (during which
time small motions may be necessary in maneuvering) compared to when
it is fastened to a very stiff concrete pad.

\subsubsubsection{Azimuth torque bias}
\label{bias}
One of the two azimuth motors is primarily used to drive the antenna
clockwise, while the other motor is used to drive it anti-clockwise.
To provide a torque bias, the alternate motor is driven with 10\% of
the torque in the opposite direction.  This bias minimizes backlash
and gear wear as the motor velocities vary to compensate for small
disturbances during slewing and tracking.  In the case that the
primary motor reaches its torque limit, the second motor is commanded
to reverse and join the primary motor.  The maximum acceleration has
been set such that this only happens very rarely, such as during a
particularly strong wind gust.  The servo software also supports the
option of single azimuth motor operation, in which case that motor
receives the torque command directly and there is no torque bias.

\subsubsection{ACC software}
\label{accsoftware}
For remote operation of the antenna, two processes on the ACC are
started.  The first process (called ``track'') handles the ephemeris
information and computes new positions and tracking velocities in
altitude and azimuth coordinates once per second.  Details on the
calculations as a function of universal time and celestial coordinates
are described in \citet{Patel00}.  These positions and velocities are
transferred via a segment of shared memory to a second process (called
``servo'') which uses these data to compute and send new velocity and
acceleration requests to the SCB at a rate of 100~Hz in the form of
serial packets.  The SCB responds to each of these packets with its
status, its torque commands, and the velocities from the tachometers.
``Servo'' takes care of any error conditions in the SCB or the drive
amplifiers, shutting down in response to any dangerous errors.  It
passes 33 items of status information back to ``track'' in another
part of the same shared memory segment.  In a third part of the shared
memory segment, ``servo'' maintains a circular buffer of performance
data which is updated at 100~Hz.

\begin{figure}[hb!]
\includegraphics[width=6.5in]{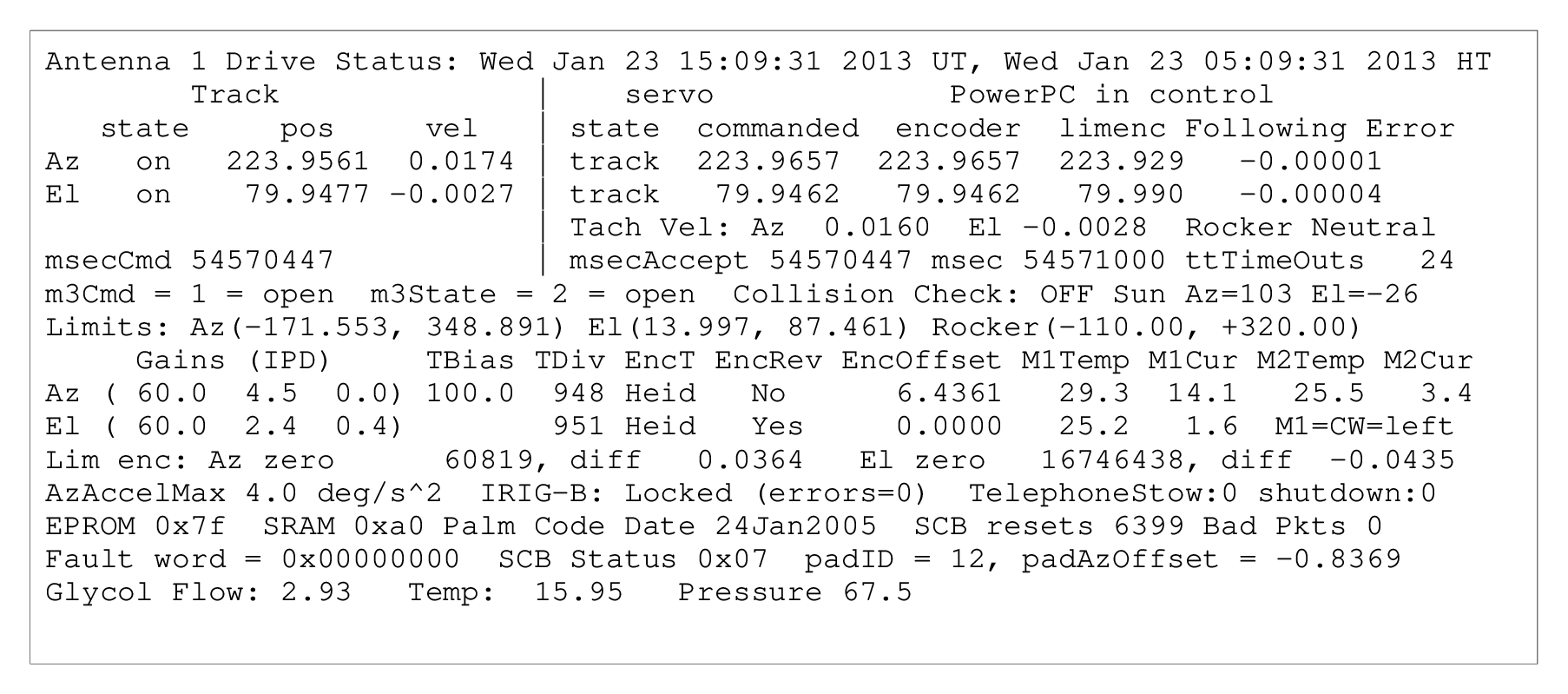} 
\caption{Snapshot of the antenna drive status display page for antenna
  1 in the text-user-interface program that monitors the complete SMA
  observatory status over dozens of pages. Any values in error state
  will appear in inverse video.  The empty bottom line of text is
  reserved for important transitory messages such as ``Emergency stop
  pressed''.
\label{curses}} 
\end{figure}

To transfer servo information to and from the antennas, there is a
fiber-optic based reflective memory system which provides a memory
segment shared between the ACC and the central control computer.
``Servo'' uses this to send messages to the array operator in case of
serious problems and to provide 78 status items for monitoring and
recording by the central system.  These include values which change
rapidly like the actual azimuth and elevation down to semi-static
items like pad ID and software version numbers.  ``Servo'' updates any
status item which has changed every second and the central system
records those which have changed once per minute.  A high level
text-user-interface (TUI) monitor program using the curses programming
library for terminal control \citep{Strang86} can display most of this
information to the operator or any other interested personnel over the
internet via a standard terminal application.  An example of the
antenna drive status display page for one antenna is shown in
Figure~\ref{curses}.

For testing, setting servo parameters, and maintenance, two additional
computer programs were written.  The first -- ''svdata'' -- reads the
performance data which ``servo'' places into the servo-track shared
memory circular buffer at 100~Hz and writes it to a text file.
Plotting this data was extremely useful in initially commissioning the
system and has remained useful for maintenance.  For example, by
recording data during complete upward and downward elevation scans, we
are able to measure the balance points of the antennas as the point
where the applied torque during the two scans are equal and opposite
(mean elevation = 78\arcdeg\/).  The second -- ''dmytrack'' --
interfaces with ``servo'' in the same way ``track'' does, but accepts
interactive operator commands for azimuth and elevation position and
velocity.

\subsubsubsection{Position servo loop parameters}

The position servo loop uses a simple proportional control algorithm
supplemented by velocity feed forward.  The gain is only 5\persec\/ so
the bandwidth is less than 1~Hz.  Fast closure after a slew is
dependent on accurate velocity feed-forward during the slew so that
the final position is near the goal.  This in turn depends on accurate
calibration of the tachometers in the SCB as discussed in
\S\ref{feedback}.

\subsubsection{Trajectory command shaping}

The nested velocity and position loops as described so far work well
as long as they stay within their linear range of operation.  In order
to avoid nonlinearities, it is essential for the position loop running
on the ACC to construct and transmit a smooth sequence of values to
the velocity loop which are within the ability of the drive system to
follow and which do not excite the mechanical resonances of the
antenna.  An algorithm designed to achieve this result is commonly
called a command preprocessor or a trajectory preprocessor
\citep{Tyler94,Gawronski02,Smith08}.  In this paper, the term command
shaper will be used.  The three main goals of the command shaper are:
(1) to avoid saturing the integrator, (2) to limit the bandwidth
required to follow the command so that no component excites a
mechanical resonance, and (3) to operate within the maximum
capabilities of the system (with a small amount of headroom to allow
for minor overshoots).  For this command shaper, the Gauss error
function (erf) was chosen as originally proposed by \citet{Woody98}.
The benefit of a Gaussian shape in the time domain is that it also has
a Gaussian shape in the frequency domain; thus, it generates less
power at frequencies that could excite the structure compared to a
trapezoidal command shape of similar duration.

Similar to the algorithm described in \citet{Olberg95}, the command
shaper enters various states depending on the current position and
velocity request and the state of motion of the antenna.  This is
implemented as a state machine which also controls starting and
stopping the drive system.  During routine observations of sidereal
sources, the state machine is typically in the TRACKING state, in
which the antenna follows the slow but steady apparent motion of a
source acoss the sky.  During the TRACKING state, if the next
commanded position ever differs by more than 5\arcsec\/ from the
previous value, or the commanded velocity differs by more than
5\arcsecpersec\/, then a commanded change of source is inferred and a
move sequence is initiated.  Given the delivery rate of new positions,
this threshold will never be triggered during sidereal tracking unless
the elevation exceeds 89\arcdeg\/ elevation, which is already beyond
the elevation software limit.  

\begin{figure}[hb!]
\includegraphics[width=7.2in]{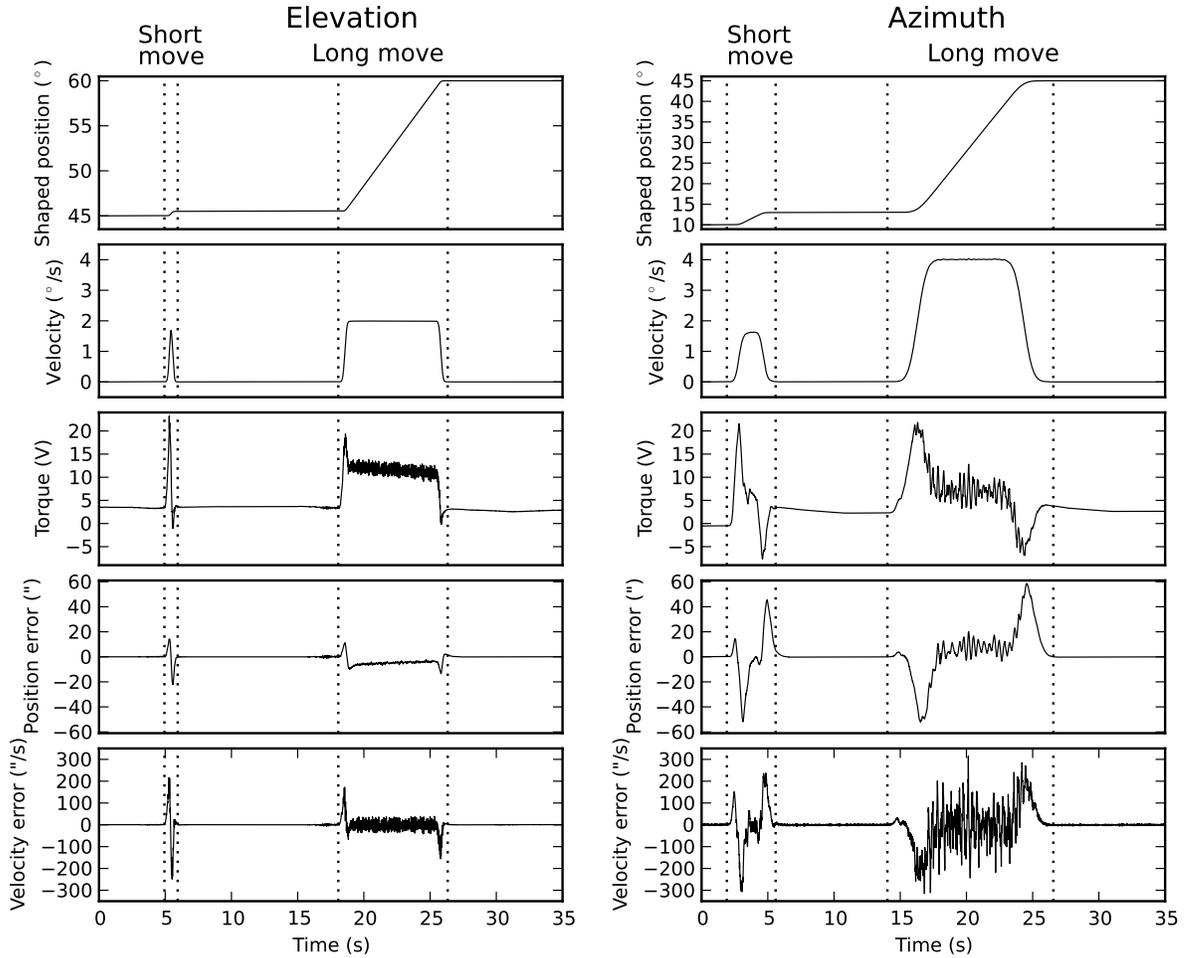} 
\caption{The response of the system to a ``short'' move followed by a
  ``long'' move in elevation (left column of plots) and azimuth (right
  column of plots).  Top row: the position command vs. time, as
  produced by the command shaper.  Second row: the velocity command.
  Third row: the requested torque.  Fourth row: the position error
  (commanded minus actual). Bottom row: the velocity error (commanded
  minus actual).  The vertical dotted lines denote the beginning and
  end of the movement segment.
\label{servo10panels}} 
\end{figure}

There are two types of move sequences implemented: ``short'' and
``long''.  A short move is one in which the position and velocity are
interpolated between their current values and the values they should
have at the projected end of the slew with a single error function.
(The erf() function has an infinite range, but it is truncated here so
that the discontinuity at each end is small but the time spent
starting or stopping is minimized.)  A long move has three parts:
acceleration to the maximum velocity, a (possible) constant velocity
part, and a deceleration to the final position and velocity.  The
acceleration part follows a truncated error function in velocity
between the present velocity and the maximum velocity while obeying
the maximum acceleration allowed.  The deceleration part is similar
and timed to join with the requested path in position and velocity.
As an example, the response of the system to a short move followed by
a long move in each axis is illustrated in Figure~\ref{servo10panels}.
When planning a move sequence, the first thing the command shaper does
is to plan a short move which stays within the velocity and
acceleration limits of the system.  If the short move takes less time
than a long move would, then it is used.  Otherwise a long move is
planned and executed.

\subsubsection{Sun avoidance}
\label{avoidance}

The SMA antennas are not designed to directly observe the Sun.  The
cabling to the subreflector can reach destructive temperatures when
the telescope is pointed near the Sun for a significant period of time
due to illumination by off-axis caustics.  To guard against accidental
movements toward the Sun, the server process on the ACC frequently
computes the solar position and the number of minutes before it will
enter the avoidance zone based on the antenna's current position. If
the number of minutes reaches zero, the Sun has entered the avoidance
zone, and the antenna is moved to avoid the Sun.  The radius of the
avoidance zone (typically 25\arcdeg\/) is read periodically from
reflective memory so it can be modified by the operator if necessary
without restarting the ``servo'' process.  If a slew command would
cross through the sun avoidance zone, but the end point is not within
it, then the path is redirected around zone.  The avoidance is active
even when the telescope is under manual control with the \palmsp.
However, the sun avoidance check is automatically disabled if the pad
identification hardware value does not match a normal pad, i.e. it
corresponds to the transporter or the indoor pad in the antenna barn.

\subsection{Automatic monitoring features}

The three servo loops (position, velocity, and motor current) perform
self-consistency checks on the state of the system and can each
disable the drive system if these checks fail.

\subsubsection{SCB}
When the drives are enabled, the following list of conditions (in
addition to normal limit switches, azimuth rocker arm deployment
check, and E-stops) will cause the SCB to command a servo shutdown:

\begin{itemize}
\item The serial packet stream from the ACC (or the \palmsp\/ if it is in control)
  fails
\item The integral of any tachometer fails to match the change in the
low resolution encoder reading on the corresponding antenna axis
\item The ADC overflows due to a large voltage from a tachometer for 5 
consecutive samples
\item A motor goes over temperature
\item The watchdog timer expires (triggering a reset of the microcontroller).
\end{itemize}

In the event that a low resolution encoder fails, it can be difficult
to move the telescope at all due to the logical interlock described in
the second item above.  Thus, an option was added to disable this
comparison between the tachometer integral and the change in the
encoder in order to deal with this rare situation.

\subsubsection{ACC}
The ``servo'' process on the ACC will command a servo shutdown if:
\begin{itemize}
\item The high resolution encoders fail to report a position
\item The position reported by the high resolution encoders 
differs substantially from the low resolution encoders (reported by the SCB)
\item The number of serial packet errors exceeds a specific threshold.
\end{itemize}

In addition, the ``servo'' process will refuse to enable the azimuth
servo if the air pressure system does not show a fault at startup time
(which it should prior to releasing the azimuth brake).  This feature
protects any person who may have manually released the brakes using
the temporary brake release button in the antenna cabin.  This button
engages an adjustable timer circuit that releases the azimuth brakes
for (up to) ten minutes to allow personnel to rotate an antenna by
manual force in case it is impossible to start the drive system.
Finally, the ACC has the ability to command a reset of the SCB
microcontroller via a digital output line on the VMEbus.

\subsubsection{Motor amplifiers}

The motor amplifiers have two electronic circuit breakers: high-speed
and low-speed.  Specifically, the amplifier will shut down immediately
if the current exceeds 75~\ampere\/, and after 15 seconds if the
current remains above 50~\ampere\/. The low-speed breaker protects the
motor in the event of binding due to mechanical overload (such as a
brake in the ``released'' state failing to fully release), or an
electrical short causing overload of the output.  These breakers have
been triggered only very infrequently in operation because the SCB
software limits the commands to the amplifier to safe values.  In
addition, an amplifier will fault and shutdown if: the motor goes over
temperature, the amplifier board goes over temperature, or any
significant failure in the supporting electronics occurs, including
power supply voltages exceeding their specification range.

\section{Portable manual interface}
\label{interface}

There are several reasons to include a manual, portable control
interface for the antenna.  In order to sample a broader range of
spatial scales on the sky, interferometers like the SMA frequently
need to have their antennas transported and redistributed across the
available pads.  To facilitate loading onto the transporter vehicle,
the antenna must be driven to zenith and to a particular azimuth.  It
is convenient to have a simple hand interface for the transporter crew
to locally operate the antenna (rather than entering cryptic commands
on a keyboard somewhere).  In the system described here, the antenna
can be driven by the SCB via the \palmsp\/ interface completely
independent of the operating state (or even the presence) of the ACC.

\subsection{\palmsp\/ computer}

\begin{figure}[h!]
\includegraphics[width=3.0in]{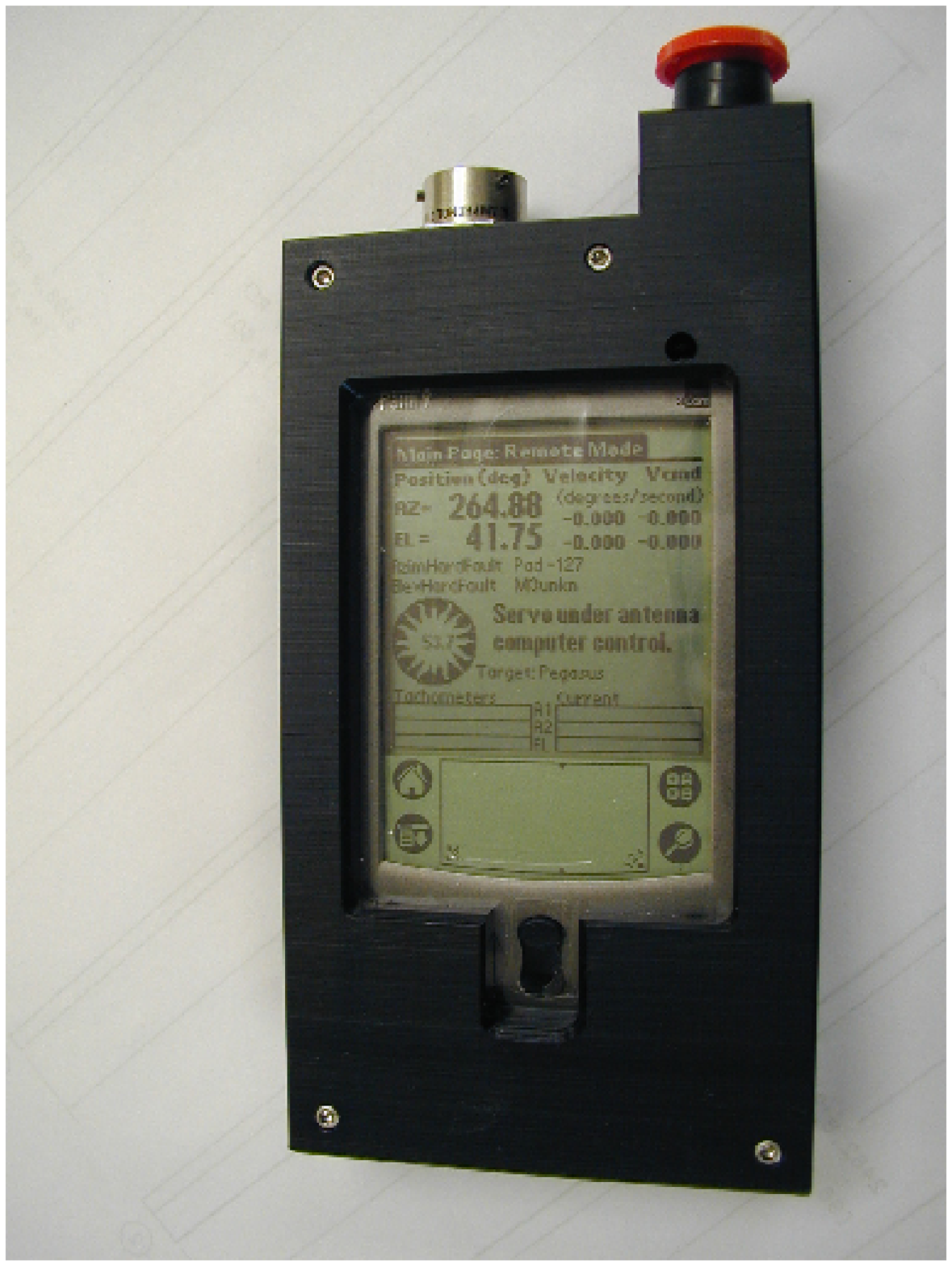} 
\includegraphics[width=3.0in]{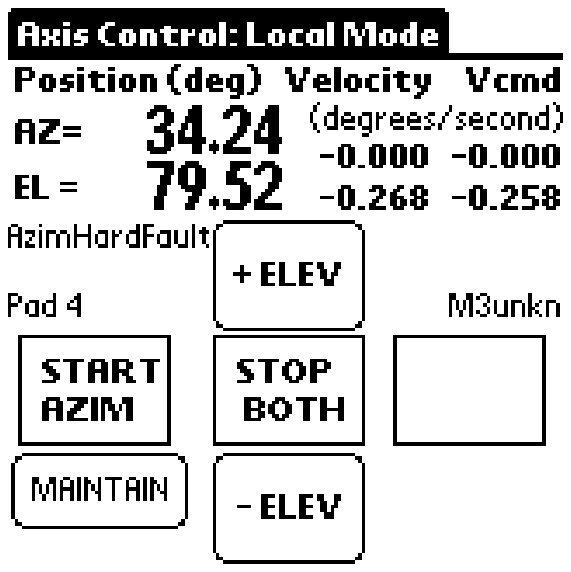} 
\caption{(a) Photograph of the \palmsp\/ hand controller embedded in
  its black Delrin \textsuperscript{\texttrademark}\/ case.  The
  circular connector at the top carries the power supply, serial
  communication lines, and the E-stop circuit wires.  It can be
  connected either inside or outside the antenna cabin. (b) An
  example display on the \palmsp\/ hand controller showing the
  elevation axis under manual control.  The $+$ELEV and $-$ELEV
  buttons will accelerate the antenna, while the MAINTAIN button will
  maintain the current velocities.  If the user releases his finger
  from the touch screen, any active axis will decelerate to zero
  within a few seconds.
\label{palm}} 
\end{figure}

The \palmsp\/ is a small, lightweight palm-sized computer package
containing a $160 \times 160$ pixel greyscale liquid crystal display
(LCD), a Motorola MC68328 microprocessor running at 20~MHz, and a
traditional RS-232 serial interface (Figure~\ref{palm}a).  Unlike
other models manufactured by Palm, it draws its power from
rechargeable, long-life lithium-ion batteries.  The LCD screen
continues to work even at low temperatures (near $0^\circ$
Fahrenheit), and a backlight is available for use in the dark.  Each
\palmsp\/ has been embedded in a protective Delrin
\textsuperscript{\texttrademark}\/ case that is comfortable to hold in
one hand, and has a replaceable transparent window to protect the
screen from the elements.  There is a hardware E-stop button mounted
onto the top of the case which is placed in series with all other
E-stop buttons on the antenna.  The wires for this button, along with
the serial data and power supply lines all run through a single cable
with a circular connector.  At the other end of the cable, the power
supply lines are split out into a separate regulated power
supply. With this power connection, the lithium-ion batteries will
remain fully charged despite the continuous drain from the serial line
circuitry.

\subsection{Manual control features}

The SCB continuously transmits status values to the \palmsp\/ which
serves as a convenient display terminal for personnel inside the
antenna cabin.  This condition is termed ``remote mode''.  The
operating program on the \palmsp\/ is written in C and has many
different pages that display the antenna position, velocity, motor
currents and temperatures, as well as encoder zero values, servo gains
and filters, and tachometer calibration settings.  More powerfully,
the \palmsp\/ program can also be used to take manual control of the
velocity servo loop away from the ACC into a ``local mode''.  If local
mode is requested while the servo is active under remote control, it
is first smoothly decelerated and disabled before control is passed.

Local mode always begins with the servo disabled.  One or both axes of
the antenna can then be activated via the touchscreen. The antenna can
then be moved up and down in elevation and/or left and right in
azimuth using a traditional cross layout of software buttons on the
touchscreen (Figure~\ref{palm}b).  These buttons increment or
decrement the velocity, or stop the antenna.  In effect, only one
button can be depressed at a time.  When neither button for an axis is
depressed, the SCB automatically slowly decelerates toward zero
velocity.  If the button is depressed again, acceleration would resume
from that point.  If the velocity is non-zero, then the ``maintain''
button appears which, if depressed, will maintain the current velocity
in both axes.  The soft stop button in the center will automatically
engage a rapid decaying velocity sequence down to zero velocity prior
to engaging the brakes.  While all of this is happening, the antenna
position and velocities are displayed and updated on the touchscreen
at about 1~Hz.  There is also a slow speed mode called ``turtle
mode'', in which the velocity steps are reduced by a large factor and
the velocity units displayed change from \degpersec\/ to
\minpersec\/. This mode is useful for fine positioning of the antenna
under manual control, for example when installing the subreflector in
the antenna barn.  The clock time on the \palmsp\/ is updated
periodically from the ACC clock time, which allows the \palmsp\/ to
display the angle between the Sun and the telescope pointing
direction, alerting the user with a flashing field when it becomes
less than the avoidance criterion (see \S\ref{avoidance}).

For safety purposes, the local mode never times out, however the servo
will be automatically disabled if an idle timeout of a few minutes is
reached.  If the \palmsp\/ is ever disconnected or otherwise fails to
communicate while in local mode, the SCB will immediately reassert
control over the antenna, entering remote mode and shutting down the
servo if active. If the antenna has been mistakenly left in local
mode, control can be usurped remotely but only by entering the default
password in the terminal control program (and only if the local user
did not enter an optional temporary password when local control was
taken).  A third mode exists called ``lockdown'', in which the user
must enter a password.  In this mode, control cannot be usurped and no
other user can use the \palmsp\/ to move the antenna, which can offer
additional protection from unaware co-workers in the event that an
E-stop padlock system is unavailable, and one needs to climb on or
under the antenna.

Further details on the operation of the \palmsp\/ can be found in
\citet{Hunter00}.  The \palmsp\/ has been an intuitive and popular
device among the daily telescope operations and maintenance crew
\citep{Christensen08}.  The original devices continue to function
well, perhaps longer than expected, and sufficient spares remain on
hand.

\subsection{Emergency stowing via telephone}

The SMA operates unattended during the early morning hours of
weekdays, controlled from a room in Cambridge or Taipei
\citep{Christensen08}.  Because weather conditions can become severe
quite quickly on Mauna Kea \citep{Cherubini09}, to provide an
additional level of safety, a fourth mode of servo operation called
``telephone mode'' was programmed.  This mode was created to allow the
operator to stow the antennas in the condition that internet
connectivity is lost, but telephone connectivity remains.  A
commercial device accepts calls, and if the correct access code is
entered, it presents a menu which allows the caller to trigger a
digital input on the SCB in one or more antennas via a multimode fiber
optic cable.  After 30 seconds of continuous activation of the digital
input, if the telescope is on an outdoor pad, the SCB switches to
telephone mode.  It closes the environmental cabin door and activates
a state machine which will start the drives (if necessary) and then
accelerate each axis to 1/4 of full speed.  For each drive, it runs at
that speed until near the stow position, when it decelerates to a stop
and turns the drive off.

\section{Performance}
\label{performance}

The ``servo'' program running on the ACC has the ability to collect
and record an extensive set of real-time data to a disk file for
testing purposes.  This facility was used to record the performance of
the servo in action.

\subsection{Tracking}

\begin{figure}[h]
\includegraphics[width=7.0in]{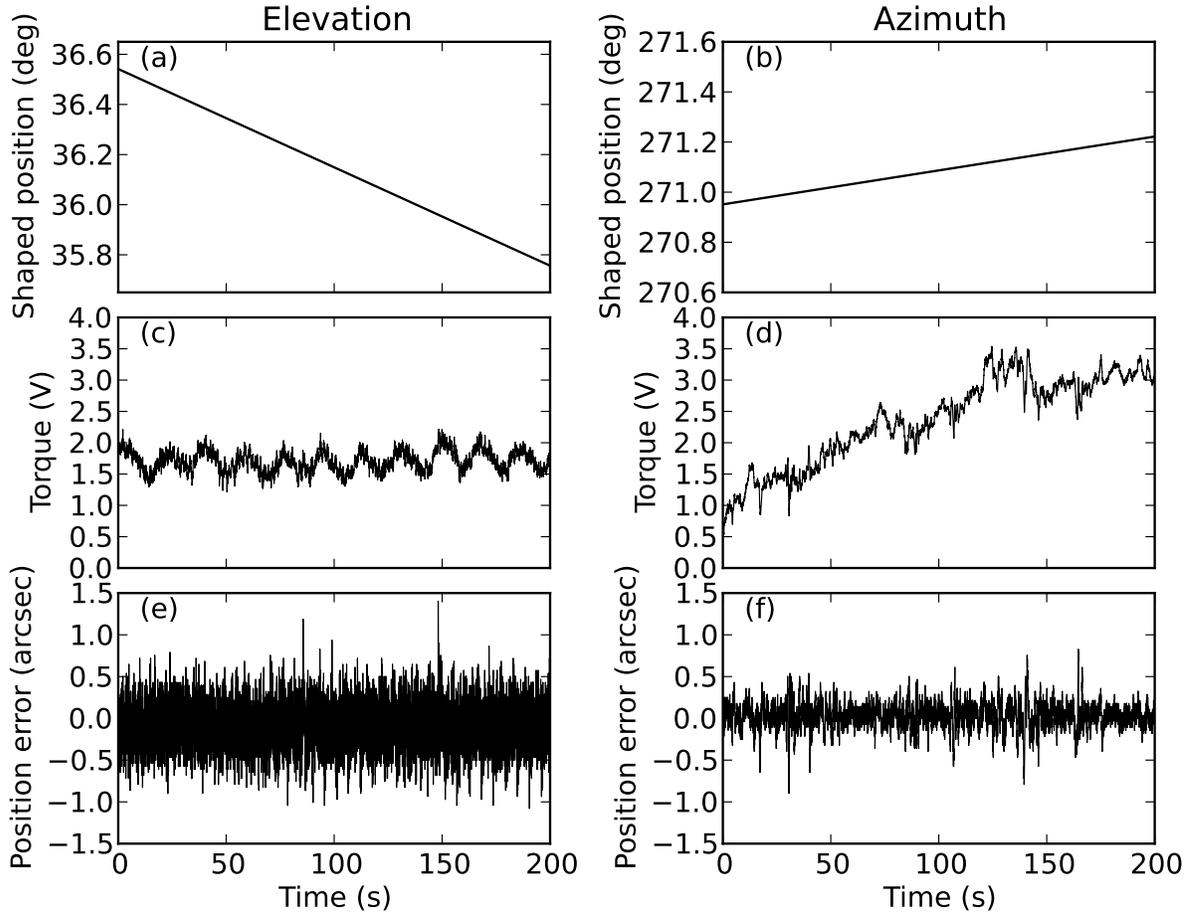} 
\caption{Servo performance while tracking a setting source.  Panels
  (a) and (b): the commanded axis position in degrees.  Panels (c) and
  (d): the torque command to the motor.  Panels (e) and (f): the
  difference between the command position and the actual position
  reported by the high-resolution encoder (arcsec).  The rms tracking
  error is 0.161\arcsec\/ in azimuth and 0.266\arcsec\/ in elevation.
\label{tracking}}
\end{figure}

The typical two-dimensional rms tracking error of the antennas in calm
conditions is $\sim 0.3$\arcsec\/ rms.  An example of data acquired
from antenna 2 on 2013 January 9, nearly ten years after the
inauguration of the SMA, is shown in Figure~\ref{tracking}.  The
windspeed was only 2.2~m\persec\/. Over a 200 second period, the rms
error in azimuth is 0.161\arcsec\/ and in elevation is 0.266\arcsec,
with mean errors of $-0.10$\arcsec\/ and +0.03\arcsec\/, respectively.
The mean total error angle (on the sky) is 0.25\arcsec\/, and the
fraction of time that this error exceeds the specification of
0.7\arcsec\/ is only 1.2\%, with a maximum excursion of 1.4\arcsec\/.
The tracking errors remain low for typical wind speeds of
$<10$~m\persec\/ but can reach 1-2\arcsec\/ for high wind speeds of
15-20~m\persec\/. The antennas are stowed for safety by the operator
when the wind speeds reach 30~m\persec\/.

Some radio telescopes suffer problems at very low tracking rates,
which occurs when the sign of the tracking direction changes.  This
will occur in elevation at meridian transit for all sources, and in
azimuth at two specific points along the track of sources that transit
between the zenith and the pole.  The performance in these conditions
has been tested on both axes, decelerating smoothly from
+0.15\arcsecpersec\/ to $-0.30$\arcsecpersec\/ over a 10 minute period
and the tracking performance is excellent: the rms is 0.11\arcsec\/ in
elevation and 0.15\arcsec\/ in azimuth.  At the lowest elevations
($\leq 15$\arcdeg\/), the elevation tracking on some antennas
currently degrades due to mechanical effects (stiction), likely caused
by the combination of a small misalignment of the ballscrew axis with
respect to the elevation axis and the elevation balance point being at
much higher elevation.

\subsection{Slew rate, acquisition time, and fast switching}

\begin{figure}[h]
\includegraphics[angle=-90,width=6.6in]{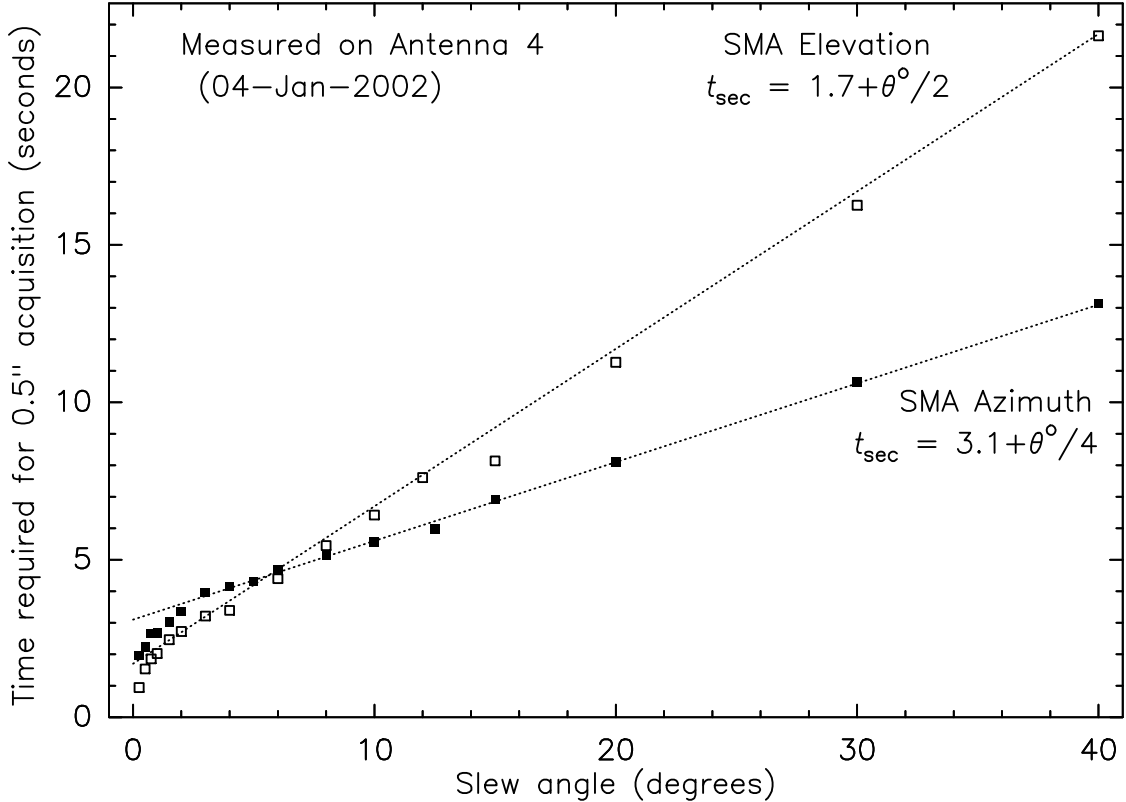} 
\caption{Measurements of the source acquisition time in seconds as a
  function of the slew angle in degrees along each axis. Filled points
  are for azimuth slews, and open points are for elevation slews.  The
  lines are approximate fits to the large angle data
  points. \label{azelslew}}
\end{figure}

Historical measurements of the time required for the antenna to slew
to and acquire a new source are presented in Figure~\ref{azelslew}.
For slews greater than $\sim2$\arcdeg\/, the total time can be
parametrized independently for the two axes in the form of a constant
plus the slew angle divided by the maximum slew velocity.  The
constants are 1.7~s in elevation and 3.1~s in azimuth.

Although standard SMA observations do not typically use fast
switching, the performance of the SMA servo has been recently measured
when switching 5\arcdeg\/ in azimuth and 2.5\arcdeg\/ in elevation
from the science target to the gain calibrator every 20~seconds.  The
SMA servo achieves a 70\% duty cycle, where the duty cycle is defined
as the time spent integrating on the target or the calibrator (with a
total tracking error less than 0.7\arcsec\/) versus the total time.
Enforcing a tighter limit of 0.3\arcsec\/, the duty cycle is still
65\%.  For these two cases, the mean total tracking errors during the
intervals after a source has been acquired are 0.175\arcsec\/ and
0.155\arcsec\/, respectively.

\subsection{Maintenance}

The servo system requires very little maintenance.  To demonstrate,
the maintenance reports stored in the SMA engineering logs accessible
from the SMA Observer Center \citep{Katz08,Petitpas10} were reviewed.
In over 10 years of deployment, only two chips have failed in service
on an SCB.  One was a digital output driver, and the other was one of
the digital input optical couplers.  Several working spare boards are
available for quick installation when such repairs are necessary.  In
addition, in one antenna, both the charger for one of the \palmsp\/
units and the power supply driving the limit switch circuitry needed
to be replaced in 2005.  There are occasional problems with the
connections of the 14 short fiber optic cables that connect the
commercial parent and daughter boards on the motor amplifiers.  In
normal operations, unexpected shutdowns of the servo are very rare,
and are usually an indicator of slowly deteriorating mechanical
conditions, such as in the quality of the mesh of the azimuth motors
to the bull gear or in the lubrication of the elevation ballscrew.

\section{Summary}

The digital servo control system for the SMA antennas was designed and
developed in the SAO submillimeter receiver lab in 1998-1999 and
commissioned on an SMA prototype antenna in late 1999 and early 2000.
Many improvements to the software were added over the subsequent two
years following the growing deployment of production antennas to Mauna
Kea.  The EPROM firmware of the SCB has remained unchanged since
November 2002.  The system has been reliable, provided the necessary
flexibility for operations, and has been easy to maintain despite the
sometimes harsh conditions of the mountaintop site. The system
represents a successful example of a modern digital servo system which
can serve equally well for the routine operations of a facility
observatory like the SMA, or as a platform for the development and
testing of advanced servo control algorithms.

\acknowledgments

We thank the anonymous referee for many suggestions that improved the
manuscript.  The National Radio Astronomy Observatory is a facility of
the National Science Foundation operated under agreement by the
Associated Universities, Inc.  This research made use of NASA's
Astrophysics Data System Bibliographic Services.  The authors extend
special thanks to those of Hawaiian ancestry on whose sacred mountain
we are privileged to be guests.

\end{document}